# Epitaxial formation of ultrathin HfO$_2$ on graphene by sequential oxidation

Zhenjing Liu[1#], Qian Mao[2#], Varun Kamboj[1], Rishabh Kothari[1], Paul Miller[1], Kate Reidy[1], Adri C. T. van Duin[2], R. Jaramillo[1*], Frances M. Ross[1*]

1. Department of Materials Science and Engineering, Massachusetts Institute of Technology, Cambridge, MA, USA
2. Department of Mechanical Engineering, The Pennsylvania State University, University Park, PA, USA

*Correspondence: rjaramil@mit.edu; fmross@mit.edu
\# These authors contributed equally

**Abstract**

We demonstrate the formation of epitaxial, ultrathin hafnia (HfO$_2$) on graphene. Monoclinic hafnia (m-HfO$_2$) forms as the end of a series of sequential oxidation reactions. Starting from Hf metal grown epitaxially on graphene, oxidation leads first to an amorphous suboxide (a-HfO$_x$), then to a crystalline, hexagonal suboxide (h-HfO$_x$) in epitaxial relationship with the substrate, and finally to m-HfO$_2$ that is also epitaxial. We use scanning transmission electron microscopy to characterize the epitaxial relationships and to investigate the structure of h-HfO$_x$. We propose a series of displacive transformations that relate the different crystalline phases and are consistent with the observed epitaxial relationships with the graphene substrate. ReaxFF based reactive molecular dynamics simulations confirm our model of the oxide phase sequencing, and illustrate the role of graphene in promoting oxide crystallization. Our results suggest a way to achieve heteroepitaxial integration of high-performance, crystalline dielectrics with two dimensional (2D) semiconductors with an atomically sharp interface, which is also relevant to hafnia phase engineering.

## 1. Introduction

Ultrathin two-dimensional (2D) semiconductors are promising materials to enable continued scaling of microelectronics.[1–3] However, a major hurdle is the heterointegration of 2D semiconductors with oxide dielectrics. The oxides of most 2D semiconductors are not good dielectric materials (*e.g.*, $MoO_3$), so direct oxidation – which works so well for silicon – is not a generalizable solution. Direct growth of oxides on 2D semiconductors by established methods of atomic layer deposition (ALD) is challenging due to the low density of dangling bonds, which stymies the kinetics of ALD.[4–6] Research is underway into a variety of approaches to this problem, but there remains no solution that is clearly ready for future manufacturing.[7–10]

One intriguing approach is that of depositing ultra-thin metal films on 2D materials, followed by controlled oxidation. This could produce an ultrathin dielectric layer, or provide a seed layer for subsequent ALD.[11,12] Recent work, by ourselves and others,[13–17] illustrates that 3D metals (Au, Ti, and Nb) can be deposited epitaxially on 2D materials via quasi-van der Waals heteroepitaxy, provided that the 2D surfaces are sufficiently clean and that the kinetics of metal deposition is carefully controlled. Here we leverage epitaxial growth of ultrathin Hf metal films on graphene as the starting point to study controlled oxidation.

Hafnium oxidation is a complex process as there are at least eight polymorphs of $HfO_2$. At low pressure, the equilibrium phase diagram includes the monoclinic phase (m-$HfO_2$, space group P $2_1$/c) at ambient temperature, and tetragonal and cubic phases (c-$HfO_2$) at high temperature that show an enhanced dielectric constant.[14] Two orthorhombic phases appear in the phase diagram at high pressure. There also exist three ferroelectric phases that do not appear in the equilibrium phase diagram, but can be stabilized through doping and finite-size effects.[18] The diversity of oxide phases contributes to the usefulness of hafnia for resistive switching devices, which make use of ferroelectric phases and conductive suboxides.[20,21]

In this work we make and characterize yet another hafnium oxide, an oxygen-deficient (*i.e.*, suboxide) hexagonal phase (h-$HfO_x$) that is more electrically conductive than m-$HfO_2$. We describe a sequence of oxidation transformations that start from epitaxial, ultrathin hexagonal close packed Hf (hcp-Hf) metal on graphene, and proceed through an amorphous suboxide (a-$HfO_x$) and a hexagonal suboxide (h-$HfO_x$), before terminating with m-$HfO_2$. We find that all crystalline phases (hcp-Hf, h-$HfO_x$, and m-$HfO_2$) remain in epitaxial alignment with the underlying graphene. The intermediate phases may be stabilized by finite-size and substrate effects. Scanning near-field optical microscopy (SNOM) indicates that the intermediate oxide phases are more conductive than m-$HfO_2$. These experimental observations are confirmed by ReaxFF molecular dynamics (MD) results, which illustrate the coexistence of crystalline (hcp-Hf, h-HfO$_x$, m-HfO$_2$) and amorphous (a-Hf, a-HfO$_x$) phases, and suggest that graphene assists in the stabilization of the hcp-Hf and h-HfO$_x$ phases by templating crystallization.

**2. Experimental Results**

We deposit ultrathin Hf metal films (nominal thickness 0.86 nm, approximately 4 metal monolayers) on clean, few-layer (10-20 layers) graphene surfaces at elevated temperature. We allow the metal to oxidize under different conditions, and we use scanning transmission electron microscopy (STEM) to characterize the results. The details of graphene transfer, substrate cleaning, metal deposition in ultrahigh vacuum (UHV) and oxidation can be found in **Methods**.

*2.1. Phases formed during native oxidation of ultrathin epitaxial Hf metal on few-layer graphene*

To study the initial stages of native oxidation - *i.e.*, spontaneous oxidation in ambient conditions – we deposit a Hf metal thin film in UHV, and then expose it to air for approximately 15 min before imaging by aberration-corrected STEM. We use a relatively high substrate temperature of 600 °C (nominal) for film deposition, to promote the growth of large, faceted crystalline islands by suppressing nucleation density and enhancing the adatom diffusion length. We observe oriented islands with triangular symmetry on the substrate (**Fig. S1**). The average island lateral size is 190 nm and the area coverage is 38%, implying an average island height of 2.3 nm calculated from the total deposited thickness and coverage. The few-layer graphene substrate is supported on a perforated, electron-transparent SiN$_x$ membrane. There are no apparent differences between the Hf islands that form on and off the holes in the membrane, presumably because the 2D crystal is sufficiently thick that roughness of the SiN$_x$ membrane does not affect its surface.[14]

After 15 min of air exposure, the islands appear mostly unoxidized as they still show the contrast and diffraction spots expected from hcp-Hf, **Fig. 1a-c**. A comparison between atomic resolution images of the island and substrate, **Fig. 1b**, and an atomic model (space group P6$_3$/mmc) confirms that the (0001) planes of Hf are parallel to those of the graphene. In **Fig. 1c** we confirm this with a selected area electron diffraction (SAED) pattern, in which we identify two distinct sets of reflections. The inner set (light blue) corresponds to hcp-Hf, and the outer set (red) corresponds to graphene. We observe two epitaxial orientation relationships: the majority of Hf islands appear with a 30° rotation between Hf ($10\bar{1}0$) and graphene ($10\bar{1}0$), and a minority of islands appear with ($10\bar{1}0$)$_{Hf}$ || ($10\bar{1}0$)$_{graphene}$.

As well as hcp-Hf, he images also show a small amount of amorphous material visible at the island edge (**Fig. 1a**). We assume that the amorphous material is a-HfO$_x$ and that it is also present as a thin layer covering the entire islands; this is not apparent in plan-view STEM but can be seen in cross-section imaging, as discussed below. Upon close inspection, we also find two crystalline oxide phases present near the edges, although in limited quantities (**Fig. S2**). Their crystal structures and chemical compositions are discussed in detail below.

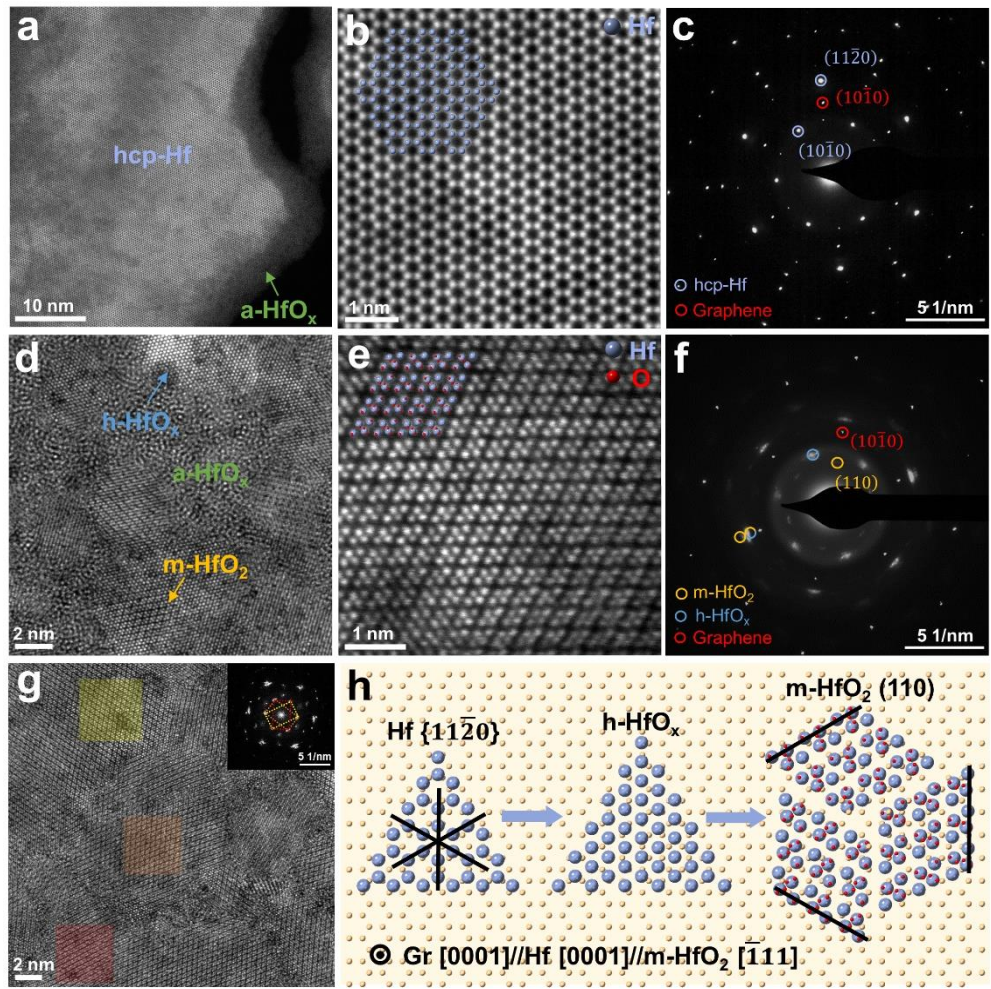

**Figure 1: Native oxidation of epitaxial Hf crystals on graphene.** (a-c) After 15 min of air exposure. (a) The appearance of a-HfO$_x$ at the island edges. (b) Crystal structure of hcp-Hf compared with atomic model as viewed along zone axis [0001]. (c) SAED pattern. (d) The co-existence of a-HfO$_x$, h-HfO$_x$ and m-HfO$_2$ during native oxidation. (e) Crystal structure of m-HfO$_2$ compared with atomic model with zone axis [$\bar{1}$11]. (f) SAED patterns collected after one month of oxidation showing disappearance of sharp hcp-Hf spots and appearance of oxide spots spread into arcs, indicating a few degrees distribution of rotational orientation in-plane. (g) Three equivalent orientations of m-HfO$_2$ domains. Inset is the corresponding FFT. (h) A summary of the epitaxy among graphene substrate, hcp-Hf and hafnium oxides.

We examined the same island after one month of native oxidation, during which the sample was stored in a vacuum desiccator under ambient temperature and sub-ambient O$_2$ and H$_2$O partial pressure conditions. We found that the Hf metal lattice had almost disappeared, replaced by a-HfO$_x$ and two crystalline oxide phases (**Fig. 1d** and **Fig. S3**), that had been barely evident after only 15 min of air exposure. One of these is identified as [$\bar{1}$11]-oriented m-HfO$_2$, by comparing its real-space atomic structure with the atomic model (**Fig. 1e**), and by comparing its fast Fourier transform (FFT) with the simulated diffraction pattern (**Fig. S4a-b**). The other crystalline phase shows a hexagonal pattern, but is inconsistent with hcp-Hf (**Fig. S4c**), and is also unlike previously-reported hafnium oxides. When visible in cross sectional images, as shown below, ABC-stacked layers can be seen. We therefore denote this phase as h-HfO$_x$. SAED (**Fig. 1f**) shows diffraction spots from both m-HfO$_2$ and h-HfO$_x$. Both are elongated into small arcs, indicating a few degrees of misorientation. Also visible is a diffuse ring that overlaps the 0.31 nm d-spacing of h-HfO$_x$, and that we assign to a-HfO$_x$. We suggest that a-HfO$_x$ may be a disordered precursor to h-HfO$_x$, with short- but not yet long-range order. We finally note that the height of the oxide islands after one month of native oxidation is 3 nm at the edges and up to 6 nm in the center, as measured by atomic force microscopy (AFM, **Fig. S5**). The edge-to-center variation comes from a similar variation in the initial metal islands, and the overall thickness increase is due to volume expansion during oxidation.

We now discuss the orientation relationship between the m-HfO$_2$ and the graphene substrate, in light of the imaging and diffraction data discussed above. The m-HfO$_2$ diffraction spots corresponding to $d_{(110)} = 0.37$ nm show six-fold symmetry (**Fig. 1f**), whereas the m-HfO$_2$ crystal structure has two-fold symmetry. The diffraction pattern can be explained by the summation of three epitaxial domain orientations, a hypothesis that is confirmed by real-space images that show all three domain types (**Fig. 1g**). In **Fig. 1h** we summarize the orientation relationships between the graphene substrate, the as-deposited hcp-Hf, and the m-HfO$_2$ that result from native oxidation.

The three oxide phases a-HfO$_x$, h-HfO$_x$, and m-HfO$_2$ remain present after an additional eight months of native oxidation (nine months in total). We conclude that, by one month of native oxidation, the process has self-limited. The typical microstructures observed after 4, 6 and 9 months of oxidation are presented in **Fig. S6**.

According to the equilibrium phase diagram, the stable oxide at room temperature and ambient pressure is m-HfO$_2$.[22] We therefore attribute the unexpected appearance of a-HfO$_x$ and h-HfO$_x$ and their stability to finite-size effects in the ultra-thin regime, in which interface effects (including epitaxy) and electrostatic equilibration (including the Mott potential) become important. Finite-size effects resulting in varying phase outcomes during metal oxidation are widely studied. Calculations predict that amorphous oxides are thermodynamically stable over the corresponding crystalline oxides in the ultra-thin limit on metals including Cr, Al, Ti and Zr; for example, the critical thickness for amorphous oxide stability on Zr (0001) is 1 nm between 298 K and 900 K.[23–25] Experimental studies of low-temperature oxidation of Zr have revealed the formation of non-stoichiometric, amorphous oxides, with thickness between 1–2 nm.[26–28] For instance, synchrotron x-ray photoelectron spectroscopy reveals that exposing Zr (0001) to pure oxygen at 10$^{-8}$ torr and 300 K for 20 min results in an amorphous oxide bi-layer, consisting of a stoichiometric oxide (ZrO$_2$, 1 nm) atop a suboxide (ZrO$_x$, 0.5 nm).[26]

*2.2. Details of the sequential phase transformations*

In this section we explore details of the sequential transformations from Hf, to a-HfO$_x$, to h-HfO$_x$, and finally to m-HfO$_2$. We observe, although with different kinetics, similar trends in both native oxidation and thermal oxidation, a process we perform at elevated temperature to drive the system beyond the native oxide configuration. In **Fig. 2** we present STEM images recorded before and after thermal oxidation, at 400 °C in air for 10 min, of a sample that had previously been stored in a desiccator for 9 months after growth. We directly observe three conversion processes: (1) a-HfO$_x$ crystalizes into h-HfO$_x$ (**Fig. 2a, d**); (2) h-HfO$_x$ transforms into m-HfO$_2$ (**Fig. 2b, e**); and (3) growth of m-HfO$_2$ at the expense of surrounding a-HfO$_x$ and h-HfO$_x$, resulting in more distinct grain boundaries between m-HfO$_2$ domains (**Fig. 2c, f**). These conversions are also confirmed from the SAED patterns, as the diffraction spots from m-HfO$_2$ intensify, while the h-HfO$_x$ spots and the a-HfO$_x$ ring decrease in intensity (**Fig. S7**).

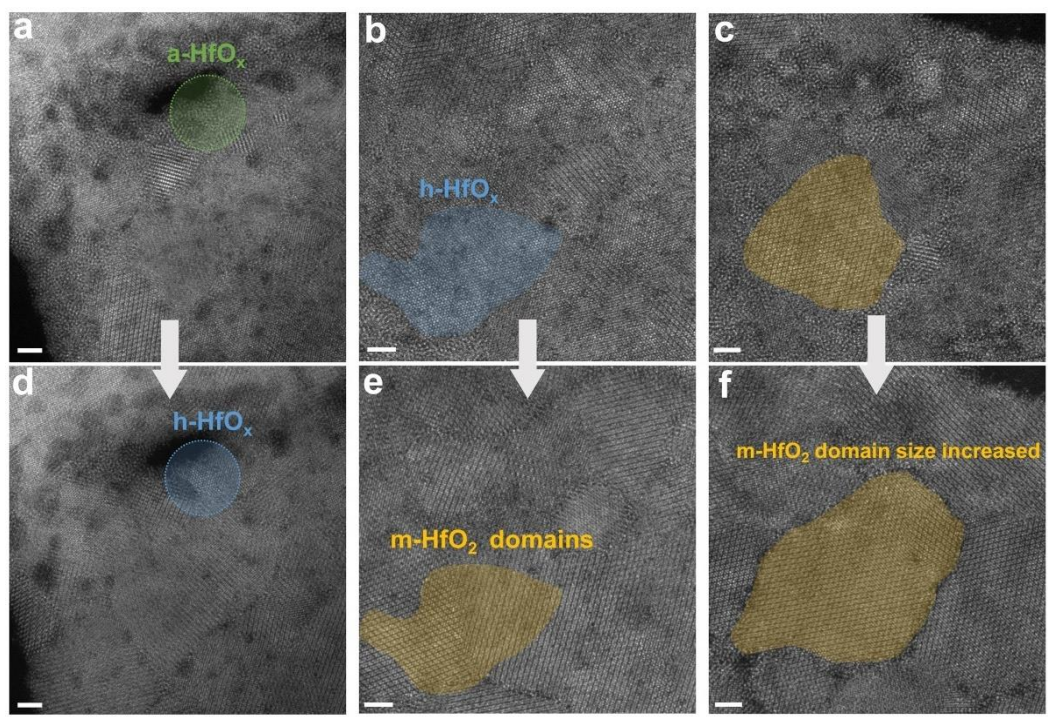

**Figure 2: Oxide phase transformations during thermal oxidation.** Image pairs show the same regions before and after thermal oxidation in air at 400 °C for 10 min. (a, d) Crystallization of a-HfO$_x$ into h-HfO$_x$. (b, e) Conversion from h-HfO$_x$ to m-HfO$_2$. (c, f) The growth of m-HfO$_2$ at the expense of surrounding h-HfO$_x$ and a-HfO$_x$. Scale bar: 2 nm.

We identify a dependence of the oxidation sequence on film thickness, by comparing two areas on the same island but with different thickness (**Fig. S8**). In the thinner area located at the island edge (**Fig. S8f**), a-HfO$_x$ remains the predominant phase. In the thicker area located at the center, (**Fig. S8e**), the crystalline phases are more prominent. This observation is consistent with the overall phase sequence, with a-HfO$_x$ forming first, followed by crystallization, with effects such as strain possibly altering the reaction kinetics.

This sequence of phase transformations can be understood by considering the starting and final crystal structures and compositions, **Fig. 3**. In **Fig. 3a** we illustrate [0001]-oriented hcp-Hf, with AB stacking along [0001] direction, and an interlayer spacing of 0.253 nm. In **Fig. 3c** we illustrate [$\bar{1}$11]-oriented m-HfO$_2$. The [$\bar{1}$11] direction is tilted by 4.8° from the ($\bar{1}$11) normal. The Hf atoms are arranged with ABC stacking, and the d-spacing of the ($\bar{1}$11) planes is 0.314 nm. The transformation from hcp-Hf to m-HfO$_2$ therefore requires a change of stacking sequence from AB to ABC, a 4.8° tilt of the Hf layers, and a 24.1% interlayer expansion of the Hf sublattice. We suggest that this substantial transformation is broken down into smaller steps by virtue of the intermediate phases. In particular, the h-HfO$_x$ structure that we infer from our images (**Fig. 3b**, top; **Fig. S9a**) shows ABC stacking. Furthermore, the in-plane expansion of the Hf sublattice in h-HfO$_x$, relative to hcp-Hf, is similar to that of m-HfO$_2$ (**Table 1**); and the interlayer spacing of h-HfO$_x$ is measured as 0.267 nm (**Table 1**), while the spacing of the ABC-stacked (111) planes in c-HfO$_2$ is 0.290 nm. We define a supercell in a single layer Hf sublattice and compare the corresponding in-plane lattice parameters ($a$, $b$, and $\theta$) in **Fig. 3d**. As listed in **Table 1**, from hcp-Hf to h-HfO$_x$ there is a 7.6% expansion of the Hf sublattice. Therefore, for reasons of symmetry and of geometry, we propose that ABC-stacked h-HfO$_x$ is an immediate precursor to m-HfO$_2$, consistent with observations in **Fig. 2b** and **e**. The final conversion to m-HfO$_2$ requires a martensitic transformation. This must be coupled to in-diffusion of oxygen, since electron energy loss spectroscopy (EELS, **Fig. S10-11**) shows that a-HfO$_x$ and h-HfO$_x$ are both sub-oxides, with oxygen content lower than that of m-HfO$_2$, consistent with previous XPS studies of Hf oxidation.[29–32] In-diffusion of oxygen for the final step of the transformation to m-HfO$_2$ is analogous to what has been described in literature for Zr oxidation.[33]

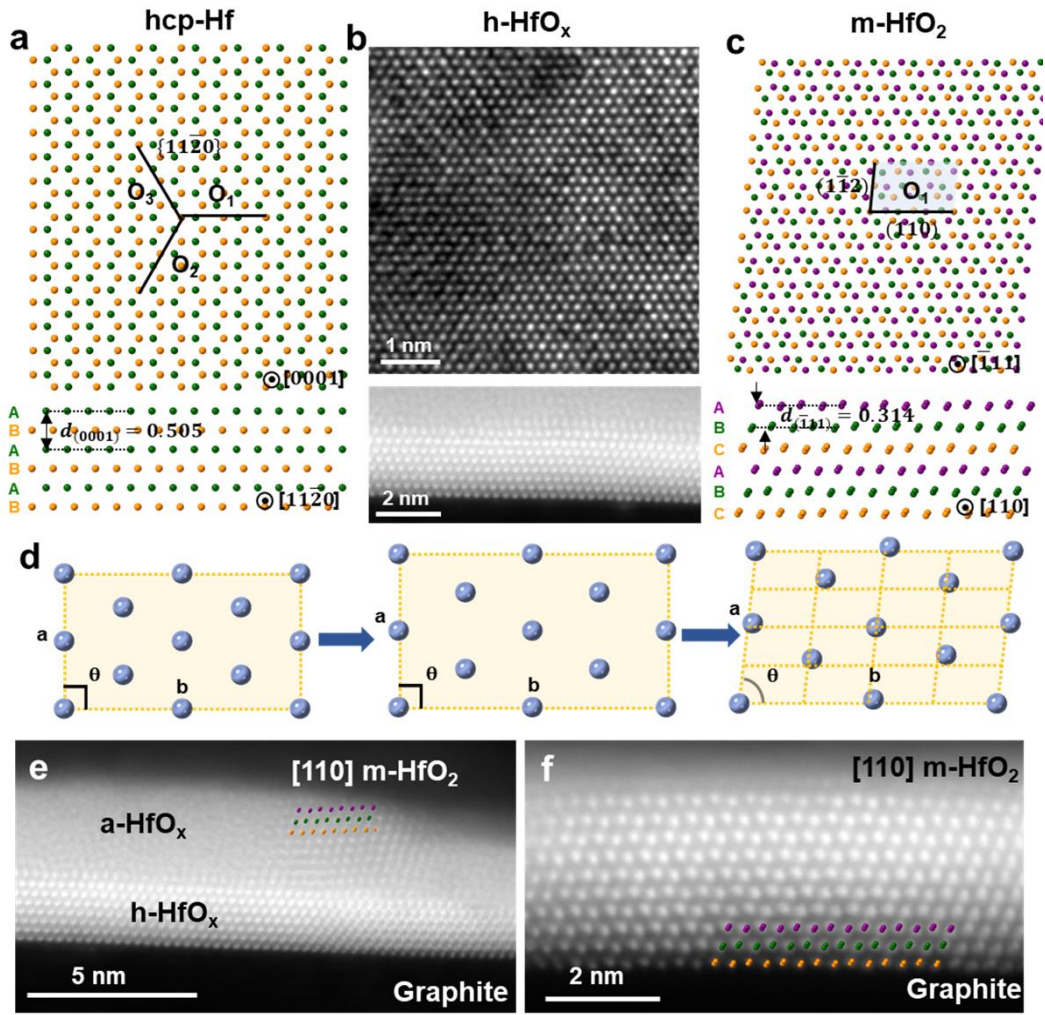

**Figure 3: Phase transformation model and discussion of the structure of h-HfO$_x$.** (a) Atomic arrangement of [0001]-oriented hcp-Hf viewed along [0001] and [11$\bar{2}$0]. Three symmetry-equivalent directions for further distortions in hcp-Hf are indicated as O1, O2 and O3. (b) HAADF-STEM images showing the h-HfO$_x$ structure, in plan view and in side-view, imaged in a lamella cut from a sample after 17 days of native oxidation using focused ion beam preparation. (c) Atomic arrangement of [$\bar{1}$11]-oriented m-HfO$_2$ along [$\bar{1}$11] and [110]. (d) In-plane transformation of single layer Hf sublattice from hcp-Hf to m-HfO$_2$ via h-HfO$_x$; $a$, $b$, and $\theta$ label the in-plane supercell lattice parameters of the Hf sublattice. (e) Cross-sectional image showing the a-HfO$_x$, h-HfO$_x$ and m-HfO$_2$ on graphite. (f) Cross-sectional image showing an atomically sharp interface between m-HfO$_2$ and graphite.

**Table 1: Lattice properties of hcp-Hf and oxide phases from experimental observations and simulations.** The columns shaded in gold, green and blue are from Materials Project, experimental observations and MD simulations respectively.

| Phase | a (nm) | b (nm) | θ (°) | Interlayer spacing (nm) |
|---|---|---|---|---|
| *Experimental results* | | | | |
| hcp-Hf | 0.673 | 1.165 | 90 | 0.253 |
| h-HfO$_x$ | 0.72±0.02 | 1.26±0.03 | 89.9±1.0 | 0.27±0.01 |
| m-HfO$_2$ | 0.75±0.01 | 1.23±0.03 | 85.4±0.9 | 0.31±0.01 |
| *Simulation results* | | | | |
| hcp-Hf ReaxFF minimized | 0.639 | 1.106 | 90.000 | 0.252 |
| hcp-Hf at 900 K | 0.639 ± 0.018 | 1.100 ± 0.020 | 89.988 ± 1.516 | 0.260 ± 0.008 |
| m-HfO$_2$ ReaxFF minimized | 0.731 | 1.271 | 82.900 | 0.319 |
| h-HfO$_x$ (Hf/Gr/50 O$_2$ at 900 K) | 0.633 ± 0.016 | 1.086 ± 0.013 | 89.983 ± 1.619 | 0.260 ± 0.011 (AB) <br> 0.262 ± 0.024 (ABC) |
| h-HfO$_x$ (Hf/Gr/100 O$_2$ at 900 K) | 0.649 ± 0.019 | 1.097 ± 0.018 | 89.977 ± 1.963 | 0.256 ± 0.015 |
| m-HfO$_2$ (Hf/Gr/3072 O$_2$ at 900 K) | 0.716 ± 0.038 | 1.090 ± 0.099 | 83.513 ± 6.112 | 0.257 ± 0.013 |

The presence of ABC stacking in the h-HfO$_x$ phase should cause the disappearance of diffraction spots with *d*-spacing of 0.31 nm, as these are related to the $\{10\bar{1}0\}$ planes ($d_{\{10\bar{1}0\}}$ = 0.28 nm) of hcp-Hf (**Fig. S9b**). However, we do observe diffraction spots with $d$ = 0.31 nm in samples after both native and thermal oxidation. We infer that the change in stacking sequence from AB to ABC is incomplete, and that the h-HfO$_x$ phase has a mixture of AB and ABC stacking. Cross sectional imaging (**Fig. 3b**, bottom) indeed shows that both AB and ABC stacking can be present. This is consistent with the hypothesis of a transformation from AB to ABC stacking, as proposed above, and with the diffraction data. The presence of ABC stacking implies a structure that is distinct from hexagonal suboxides reported previously (Hf$_6$O and HfO$_{0.7}$) that are thought to be ordered phases of oxygen dissolved into hcp-Hf.[34–36] Since the h-HfO$_x$ phase shares ABC stacking with c-HfO$_2$, but differs from the cubic phase in stoichiometry and in its lattice constants, this phase may be thought of as a structurally-distorted form of the cubic ZrO$_2$ phase, which forms during oxidation of Zr with (111) parallel to Zr (0001), and has a high concentration of O vacancies (45% on the oxygen sublattice).[37]

The h-HfO$_x$ phase identified here may be inextricably linked to its processing history and nanoscale boundary conditions. For all oxide islands that we examined in cross section, we observed that h-HfO$_x$ when present was at the graphene/oxide interface, amounting to at least part (but never all) of the interface area. We also observed that h-HfO$_x$ is always covered by a-HfO$_x$. We also observe the formation of oriented m-HfO$_2$ atop h-HfO$_x$ (**Fig. 3e**). Combined with the observed transformation sequences (**Fig. 2**), this suggests that h-HfO$_x$ forms at the interface between graphene and a-HfO$_x$, that its stability may be enhanced by the graphene lattice, and that it templates the eventual epitaxy between m-HfO$_2$ and graphene that forms during thermal oxidation (**Fig. 3f**).

### 2.3. Making continuous, ultra-thin hafnium oxide films on graphene

The above results demonstrate that ultra-thin Hf films deposited on graphene can begin as epitaxial metal-on-graphene, and end as epitaxial m-HfO$_2$-on-graphene, with appropriate processing history. To improve control over this transformation, it is helpful to separate more clearly the steps of metal film deposition and oxide formation. To do so, we carried out deposition of an ultra-thin Hf metal film (nominal thickness 1.7 nm, approximately 8 metal monolayers) on graphene at room temperature (RT), followed by native oxidation. It is known that refractory metals can deposit epitaxially on graphene if the graphene surface is sufficiently clean,[13,17,38,39] and we obtained epitaxy by using a pre-clean involving resistive sample heating in UHV (**Fig. 4a, S12**). Compared to the high-temperature (HT) deposition (nominally 600 °C) that was described above, RT deposition produces higher nucleation density due to lower Hf adatom diffusion length. The higher-resolution image in **Fig. 4b** shows the presence of Hf even in the darker boundary regions between islands. As a result, whereas HT deposition results in isolated and relatively thick Hf islands, in RT deposition we achieve a continuous film, without observable pinholes, and the height distribution has a FWHM of 0.66 nm (**Fig. S13** and the insert of **Fig. 4a**).

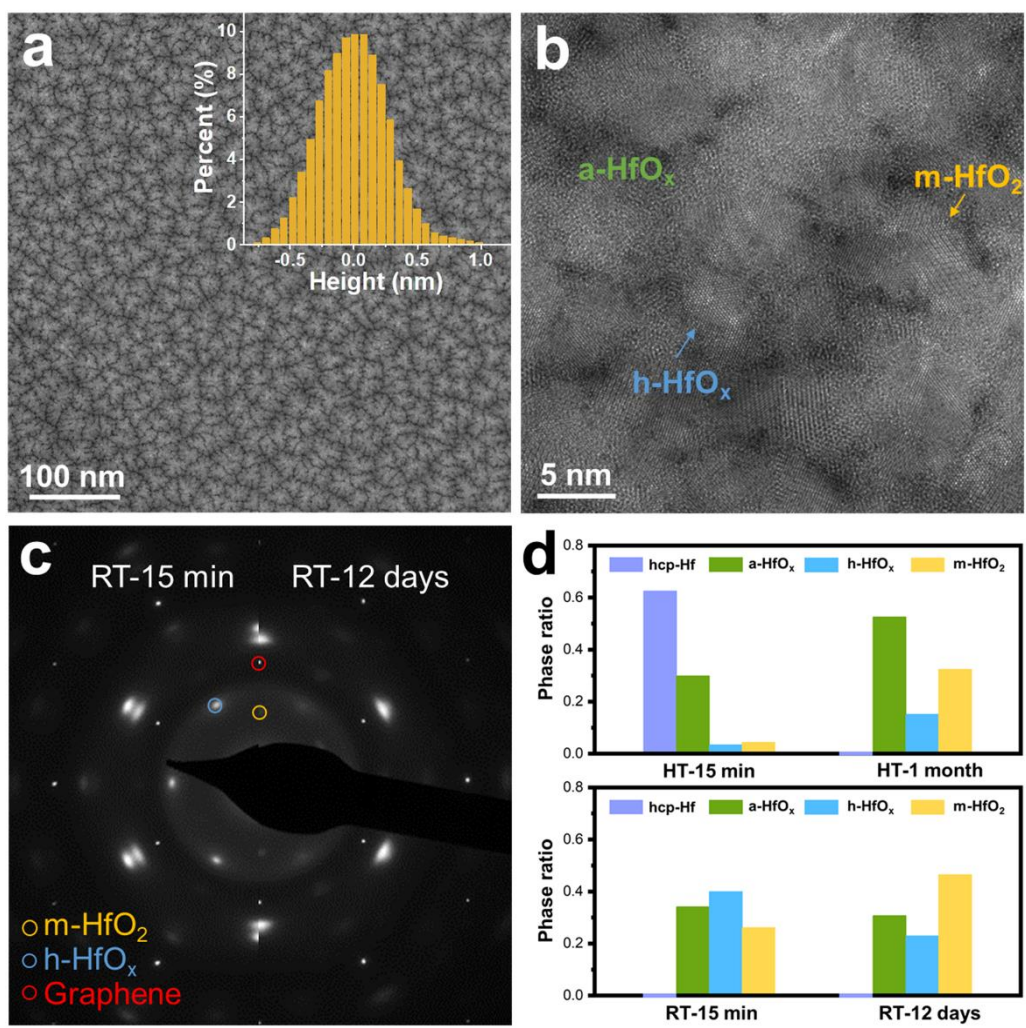

**Figure 4: Continuous and ultra-thin hafnium oxide thin film grown on graphene via room temperature deposition and native oxidation.** (a) Low- and (b) high-magnification images showing the formation of continuous $HfO_x$ film fully covering graphene substrate after 15 minutes of native oxidation. The insert in a is the height distribution histogram (relative to average height) of this

After native oxidation we observe the same a-HfO$_x$, h-HfO$_x$ and m-HfO$_2$ oxide phases in the RT film that were described above in the HT film, but the transformation kinetics are faster for the RT film. After 15 minutes of native oxidation, all three oxide phases are visible, and the hcp-Hf phase has disappeared (**Fig. 4b**), and after 12 days the m-HfO$_2$ phase is predominant (**Fig. 4b-c**). In contrast, in the HT sample, the hcp-Hf metal is slower to oxidize, and the m-HfO$_2$ phase does not predominate even after 1 month (**Fig. 4d**). The RT and HT films have comparable nominal starting Hf thickness (approximately 2 nm), but the island lateral sizes are quite different: 190 nm for the HT film, and 25 nm for RT film. We suggest that the faster oxidation kinetics in the RT film result from this difference in island size and exposure of surfaces other than the flat Hf (0001).

### *2.4. Comparing the electrical properties of ultra-thin oxides of Hf*

The usefulness of Hf oxides is connected to their electrical properties. To compare the electrical properties of the ultra-thin oxides that we observe here, we use scattering scanning near-field optical microscopy (s-SNOM) with mid-infrared (mid-IR, 10 $\mu$m) illumination. Mid-IR s-SNOM enables semi-quantitative measurements of electrical conductivity with spatial resolution below 20 nm. We performed measurements on a sample with a HT Hf metal film after undergoing thermal oxidation in at 400 °C for 10 min. In **Fig. 5a-b** we present images of topography and second harmonic scattering amplitude $S_2(f)$ (see **Methods**). The topography shows the expected dendritic oxide islands that were also observed in the STEM and AFM data (**Figs. S1, S3, S5,** and **S8**). The $S_2(f)$ data shows clear contrast between the oxide islands and the conductive graphene substrate. The $S_2(f)$ data also show an inverse correlation with island thickness, reflecting weaker mid-IR scattering where the islands are thickest, as illustrated by a line scan at position #1 (**Fig. 5c**). We also observe significant deviations from this trend in many line scans, with local maxima in $S_2(f)$ indicating more conductive regions within the oxide islands, for example in the line scans at positions #2 and 3 (**Fig. 5c**).

Based on the STEM analysis above, we expect that the a-HfO$_x$, h-HfO$_x$, and m-HfO$_2$ phases coexist in this sample, with the suboxides being more prominent along the dendritic arms of the islands where the initial Hf was thinner. We therefore conclude that the locally enhanced $S_2(f)$ indicates that the suboxides are more electrically conductive than m-HfO$_2$, consistent with previous experiments and theoretical predictions.[21,34,40–42]

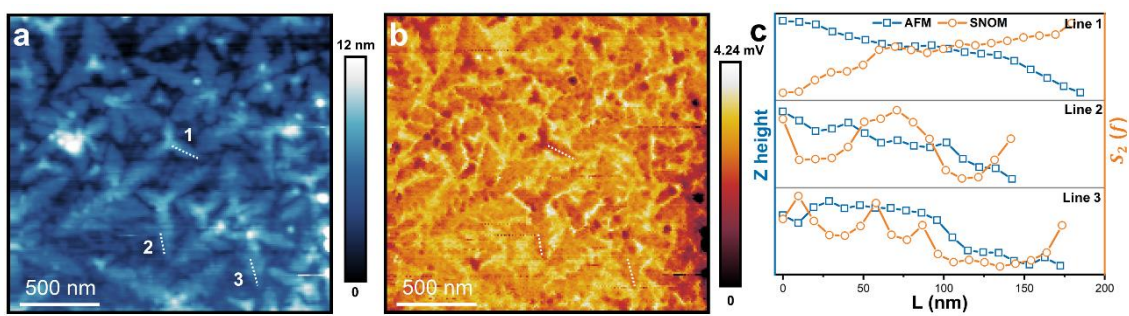

**Figure 5: Local conductivity mapping of hafnium oxides by mid-IR s-SNOM.** AFM (a) and s-SNOM amplitude (b) images of hafnium oxides on graphene after thermal oxidation. (c) Line scan profiles corresponding to the positions indicated in panels (a) and (b).

## 3. Computational modelling of oxidation pathways

We use MD simulations to study the kinetics of oxidation on length and time scales that are inaccessible to direct experimentation, yet can probe the outcomes of the oxidation processes that we described above. In **Fig. 6a-b** we illustrate the model considered in our ReaxFF MD simulations: a thin Hf metal slab on a 4-layer graphene substrate, surrounded by a variable number of oxygen molecules. The criteria for identifying distinct phases within the MD data are described in **Sec. S2.2**.

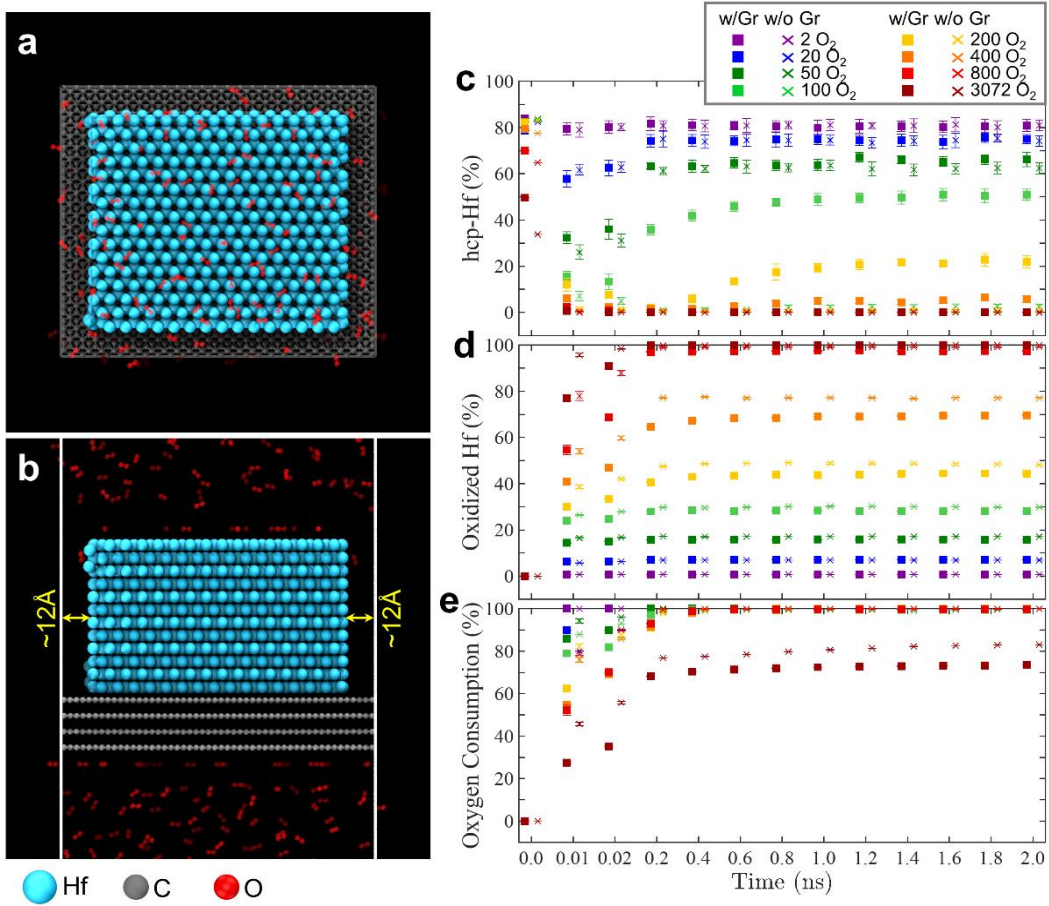

**Figure 6: The MD model and thermal oxidation behavior of Hf computed by ReaxFF.** (a) Top and (b) front views of the MD model showing an Hf slab on few-layer graphene in an $O_2$ environment. (c), (d), and (e) illustrate the percentages of hcp-Hf (Hf-Hf coordination numbers falling within 2-12), oxidized Hf, and oxygen consumption during the thermal oxidation, with and without graphene, at 900 K for 2.0 ns. Solid square symbols represent cases with graphene, and cross symbols represent cases without graphene. Each data point is an average of 11 samples, and error bars indicate their standard deviations.

*3.1. Thermal oxidation kinetics of epitaxial Hf metal on few-layer graphene*

In **Fig. 6c-e**, we present the evolution in time of a thin Hf slab during oxidation at 900 K over 2.0 ns, comparing cases with and without graphene, and for varying oxygen content. With sufficient oxygen present, the hcp-Hf metal phase disappears entirely by the end of the simulation, as expected. At intermediate oxygen content, we observe an interesting rebound of the hcp-Hf metal phase fraction (**Fig. 6c**). This is due to the formation, on very short timescales, of amorphous Hf during a period of rapid dissociation of $O_2$ molecules and Hf-O bond formation. These chemical transformations are net exothermic and therefore generate transient local heating well above the 900 K thermostat temperature (**Fig. S15**), which seems to result in the formation of a-Hf. When there is insufficient oxygen present to fully oxidize the slab, these regions of a-Hf reform to hcp-Hf within the simulation time window. In **Fig. S17** we visualize this process in greater detail. We observe the re-emergence of hcp-Hf between time $t = 1 - 3000$ ps for the case of 50 $O_2$ molecules – although noting that this regime of controllable, ultra-low oxygen concentration is experimentally inaccessible.

We note that the presence of graphene stabilizes hcp-Hf. For instance, in the case of 200 $O_2$ molecules, the slab on graphene retains a significant hcp-Hf phase fraction, whereas in the absence of graphene, the hcp-Hf phase diminishes to nearly zero (**Fig. 6c**). Two factors may contribute to this stabilizing effect: that the graphene physically shields one side of the Hf slab from the $O_2$ atmosphere, and that the graphene crystal lattice templates [0001]-oriented hcp-Hf.

The percentage of oxidized Hf (**Fig. 6d**) and the consumption of $O_2$ gas (**Fig. 6e**) vary as expected with total $O_2$ content. For intermediate $O_2$ content, the presence of graphene slightly lowers the total extent of Hf oxidation. Since the kinetics are fairly stable after 1 ns of simulation, this is evidence for the role of graphene as an epitaxial template that stabilizes hcp-Hf, in addition to being a diffusion barrier for $O_2$ gas. The effects of the graphene substrate are further discussed in **Section 3.3**.

### *3.2. Oxide phases and phase transformations*

We now describe the sequence of oxide phases and their transformations. For cases with low $O_2$ content (2–50 $O_2$ molecules), we observe h-HfO$_x$ epitaxially aligned to the graphene substrate (**Fig. S19**). Interestingly, we observe both AB and ABC stacking patterns of h-HfO$_x$, in the case with 50 $O_2$ molecules, confirming our experimental results for this previously unreported phase. As the $O_2$ content increases (100–400 $O_2$ molecules), we observe the coexistence of crystalline (hcp-Hf, h-HfO$_x$) and amorphous phases (a-Hf, a-HfO$_x$). At much higher $O_2$ content (200–800 $O_2$ molecules), the h-HfO$_x$ phase is suppressed (**Fig. S21**). At the highest $O_2$ content (3072 $O_2$ molecules), the Hf atoms are fully oxidized, with oxygen-to-metal ratio approaching $HfO_2$.

We track the oxide phase transformation through the variation in Hf-O coordination numbers (CNs). In **Fig. 7a–d** we present the distributions of Hf-O CNs for cases with a graphene substrate and with 200–3072 $O_2$ molecules. The 5-coordinated Hf atoms start to emerge with 400 $O_2$ molecules and become extensively populated with 800 $O_2$ molecules. For the case with 3072 $O_2$ molecules, we observe the appearance of 6- to 8-coordinated Hf atoms. Hf atoms with Hf-O CNs of 6–8 are likely associated with the formation of $HfO_2$ in tetragonal, monoclinic, and cubic crystal phases, respectively.[18] The 5-coordinated Hf atoms observed in **Fig. 7c** may correspond to incipient nuclei of these fully-oxidized phases.

To better identify the $HfO_2$ phases that emerge at high oxygen content (800 and 3072 $O_2$ molecules), we present in **Fig. 7e-i** the distributions of Hf-Hf bond length, Hf-O bond length, O-O bond length, Hf-O-Hf angle, and O-Hf-O angle. We also present data from an accelerated simulation with the hybrid fbMC/MD method for the case with 3072 $O_2$ molecules, and we reference distribution curves from pristine hcp-Hf, m-$HfO_2$, t-$HfO_2$, and c-$HfO_2$ (all equilibrated at 900 K in the NpT ensemble). For the case of 800 $O_2$ molecules, none of the reference $HfO_2$ phases are unambiguously present. We observe partial overlap of the Hf-Hf and Hf-O bond length distributions with those of pristine hcp-Hf and m-$HfO_2$ (**Fig. 7e-f**), but the O-O bond length and angle distributions for this case exhibit significant deviations from the references (**Fig. 7g–i**). In contrast, the result with 3072 $O_2$ molecules shows a peak in the Hf-Hf bond length distribution that is intermediate between hcp-Hf and m-$HfO_2$ (**Fig. 7e**). The first- and second-neighbor peaks of the Hf-O bond length distributions closely align with those of m-$HfO_2$ (**Fig. 7f**), and the contours of the O-O bond length and angle distributions also align with those of m-$HfO_2$ (**Fig. 7g–i**). This suggests that m-$HfO_2$ predominates in the case with 3072 $O_2$ molecules, while coexisting with a small fraction of $HfO_x$ suboxides.

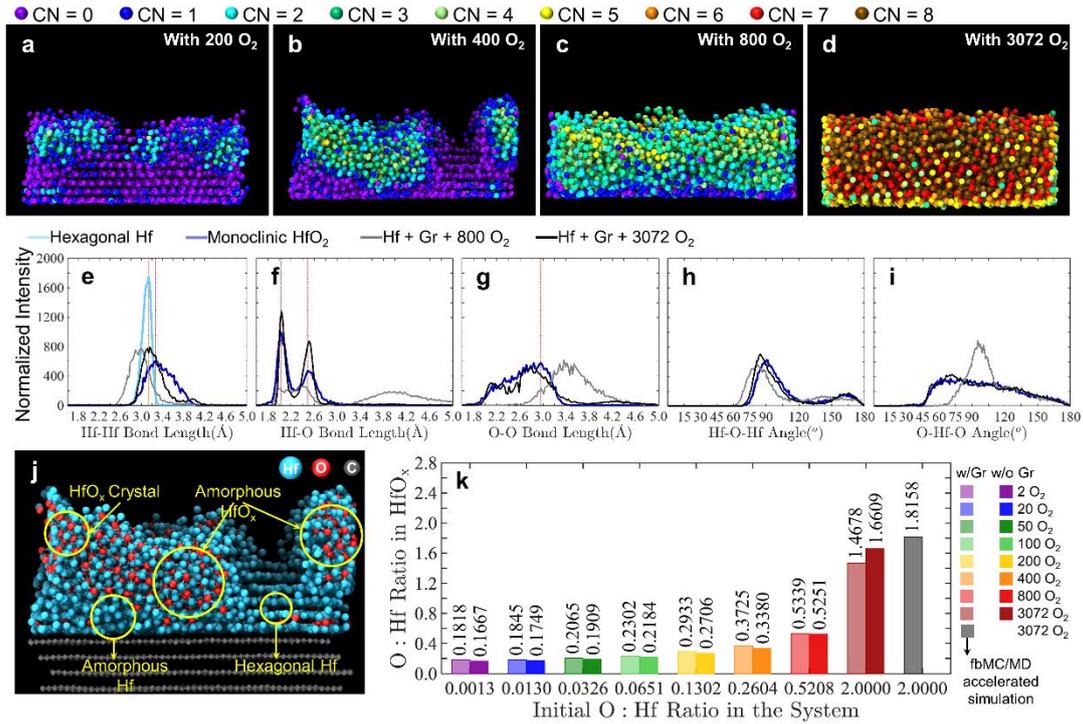

**Figure 7: Hf/O coordination numbers (CNs) and phases observed during Hf thermal oxidation.** All data are shown after annealing at 900 K for 2.0 ns. (a–d) Hf-O CNs with graphene present for 200 $O_2$, 400 $O_2$, 800 $O_2$, and 3072 $O_2$ molecules. (e–i) Bond and angle distributions for the Hf-Hf bond, Hf-O bond, O-O bond, Hf-O-Hf angle, and O-Hf-O angle, illustrating the predominant formation of the m-$HfO_2$ for the case with 3072 $O_2$ (black curves), in contrast to the case with 800 $O_2$ (gray curves). The transparent cyan and blue curves represent the characteristic bond and angle distributions for pristine hcp-Hf and m-$HfO_2$ samples equilibrated at 900 K. The vertical lines in (e) mark the 1$^{st}$ Hf-Hf bond length peaks for hcp-Hf and m-$HfO_2$, those in (f) mark the 1$^{st}$ and 2$^{nd}$ Hf-O bond length peaks for m-$HfO_2$, and the line in (g) marks the 1$^{st}$ O-O bond length peak for m-$HfO_2$ (j) Illustration showing the coexistence of amorphous and crystalline Hf and $HfO_x$ phases during the transformation sequence from hcp-Hf to m-$HfO_2$ for the case with 400 $O_2$. (k) O:Hf ratios for cases with and without graphene, and for varying $O_2$ content. The transparent black bar shows the O:Hf ratio calculated from ReaxFF fbMC/MD accelerated simulations, with the ratio being closer to 2.0, compared to those from regular ReaxFF MD. Each bar represents the average of 11 samples.

We present in **Fig. 7k** the slab-averaged O:Hf ratios at the end of the simulation time window. As expected, the average oxygen content in the oxides formed increases with total $O_2$ system content. We also see a decrease in oxidation in simulations without graphene compared to with graphene, due to the formation of smaller and more scattered $HfO_x$ clusters in the absence of graphene. For the highest $O_2$ system content studied (3072 $O_2$ molecules), the O:Hf ratio is 1.82, which is consistent with a m-$HfO_2$ predominant phase coexisting with suboxides.

In **Table 1** we present the lattice parameters found in our simulation results for slabs on graphene. For the highest $O_2$ system content studied (3072 $O_2$ molecules), the lattice parameters fall between those of hcp-Hf and m-$HfO_2$, with $a$ and $\theta$ being closer to those of m-$HfO_2$. This observation further supports the predominance of m-$HfO_2$ at high oxygen content. The computed interlayer spacing for h-$HfO_x$ is also close to the value determined by experiment.

### *3.3. Effects of the graphene substrate*

The experimental data suggests a role of the graphene substrate in the formation of the h-$HfO_x$ and m-$HfO_2$ phases. In **Fig. 8**, we present our computational analysis of the effects of the graphene substrate on the thermal oxidation of Hf. After annealing at 900 K for 2.0 ns, the pristine (*i.e.*, not-yet-oxidized) Hf slab on graphene shows a layered structure near the substrate, compared to the case without graphene (**Fig. 8b** and **e**). Additionally, the layered structure exhibits a higher fraction of the hcp-Hf phase than the case without graphene, as seen from top, front, and bottom views (**Fig. 8a–f**). We suggest that these layers promote the subsequent formation of h-$HfO_x$. This is illustrated in **Fig. 8g-l**, which shows that the presence of graphene promotes the formation of h-$HfO_x$ in cases of limited oxygen content (here, 100 $O_2$ molecules). The effects of graphene can also be seen in the spatial distributions of suboxides $HfO_x$, that form in smaller, more irregular clusters without graphene (**Fig. S24**).

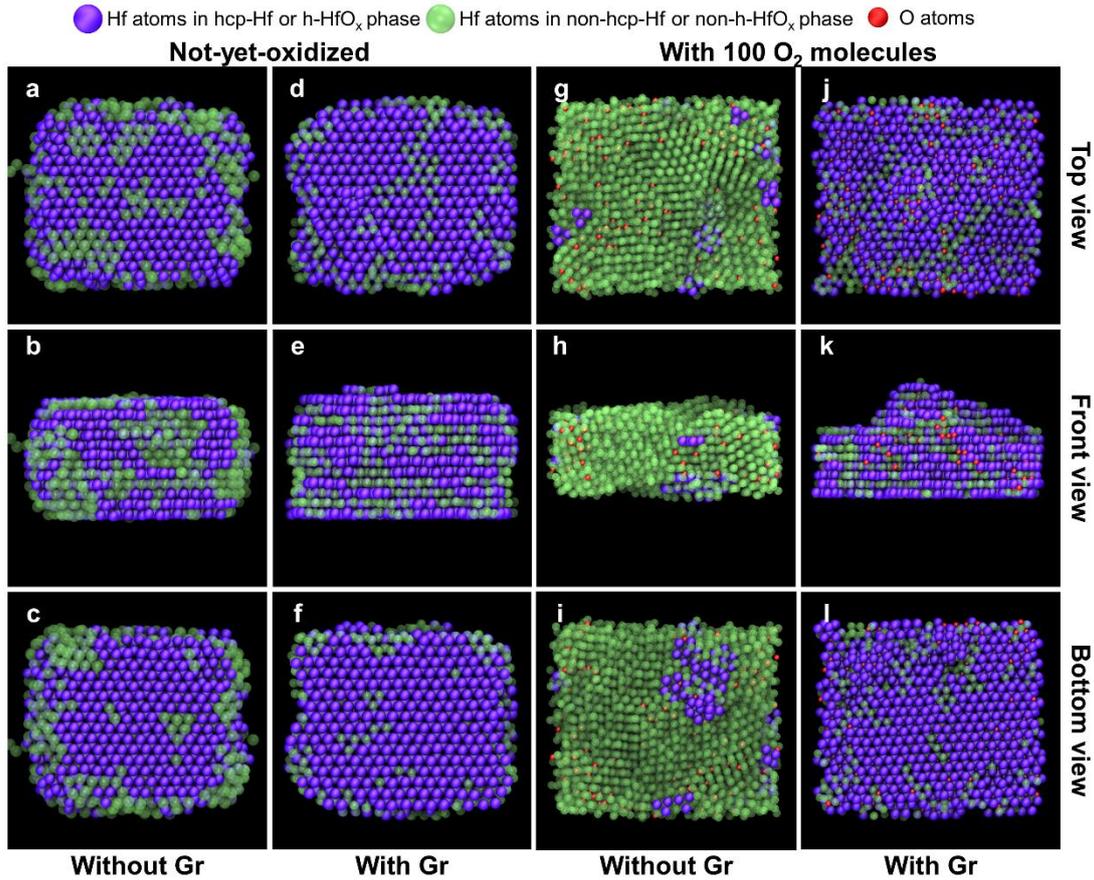

**Figure 8: Analyzing the effects of graphene substrate on Hf oxidation.** (a–c) Visualization of Hf atom distributions in hcp-Hf and non-hcp-Hf phases (top, front, and bottom views) for the not-yet-oxidized case without graphene. (d–f) Visualization of Hf atom distributions in hcp-Hf and non-hcp-Hf phases (top, front, and bottom views) for the not-yet-oxidized case with graphene. (g–i) Visualization of Hf atom distributions in hcp-Hf/h-HfO$_x$ and non-hcp-Hf/h-HfO$_x$ phases and O atoms (top, front, and bottom views) for the 100 O$_2$ case without graphene. (j–l) Visualization of Hf atom distributions in hcp-Hf/h-HfO$_x$ and non-hcp-Hf/h-HfO$_x$ phases and O atoms (top, front, and bottom views) for the 100 O$_2$ case with graphene.

In **Fig. 9**, we plot the distributions of reacted O atoms along the out-of-plane direction for cases with 50 to 3072 O$_2$ molecules, both with and without the graphene substrate. The decrease in total slab thickness, from time $t = 1$ to 100 ps, is likely related to the disappearance of amorphous phases at short times. Up to 10 ps, the difference in oxygen distribution without and with graphene is clearly related to the role of graphene as a diffusion barrier (**Fig. 9b** and **g**). At longer times, the distributions are less clearly dependent on diffusion kinetics. For $t = 100$ ps and longer, the presence of graphene leads to an enhanced oxygen content at the bottom of the slab. This supports our experimental finding that the graphene lattice promotes the formation of Hf crystalline oxides, both h-HfO$_x$ and m-HfO$_2$, through epitaxial alignment. The graphene substrate therefore contributes to directing the phase sequence. These results therefore suggest the mechanism by which we achieve atomically sharp interfaces between a 2D material and an epitaxial oxide dielectric.

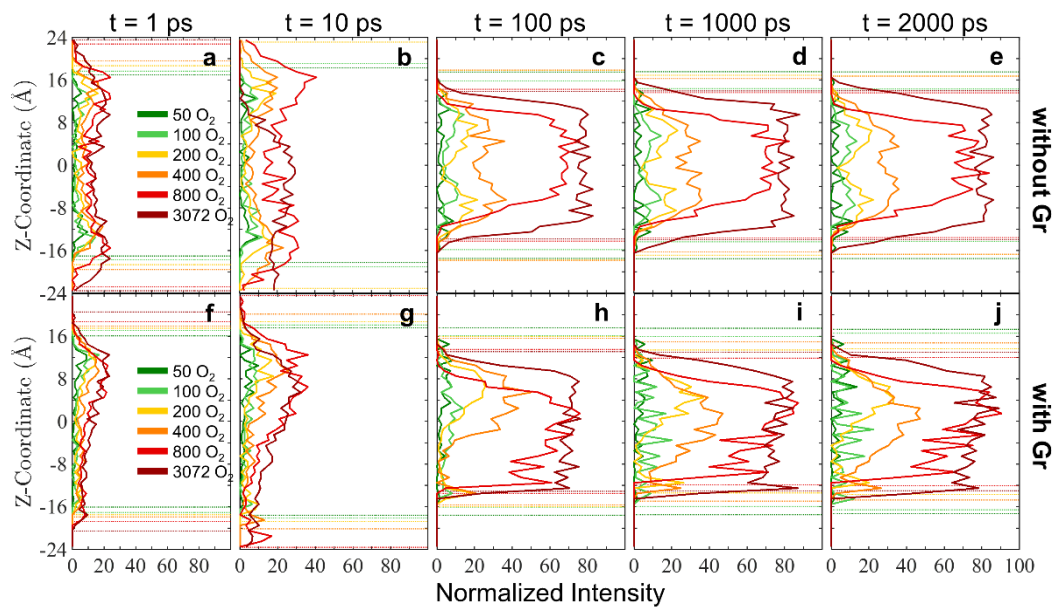

**Figure 9: Distributions of reacted oxygen along the out-of-plane direction.** (a–e) Reacted oxygen deposition along the out-of-plane direction for the Hf slab at 1 ps, 10 ps, 100 ps, 1000 ps, and 2000 ps without graphene. (f–j) Reacted oxygen deposition along the out-of-plane direction for the Hf slab at 1 ps, 10 ps, 100 ps, 1000 ps, and 2000 ps with graphene.

### 4. Conclusions

We have demonstrated the synthesis of m-HfO$_2$ on graphene with an epitaxial alignment between the two materials and an atomically sharp interface. We show that this can be achieved by the deposition of a thin Hf layer followed by controlled oxidation. Electron microscopy characterization shows that the phase sequence, beginning with ultra-thin hcp-Hf on graphene, proceeds through suboxides a-HfO$_x$ and h-HfO$_x$, before terminating with m-HfO$_2$. All crystalline phases (hcp-Hf, h-HfO$_x$, and m-HfO$_2$) form epitaxially on the substrate. The sequence can be modeled as a series of displacive transformations of the Hf lattice, coupled to oxygen in-diffusion. The h-HfO$_x$ intermediate phase appears to be similar to c-HfO$_2$ with a high concentration of stacking faults and oxygen vacancies. The formation of this h-HfO$_x$ phase is likely to be tied to the process and boundary conditions employed here, including gentle oxidation and the presence of the graphene substrate.

We also demonstrate the epitaxial deposition of a continuous Hf film on graphene using low temperature deposition on a clean substrate. This therefore constitutes a low-temperature processing pathway to form an ultra-thin (approximately 2 nm), conformal, epitaxial HfO$_2$ film on graphene.

ReaxFF based molecular dynamics simulations are consistent with the experimental results. These simulations illustrate the roles of kinetics, limited oxygen pressure, and the graphene substrate in promoting the formation of the crystalline h-HfO$_x$ and m-HfO$_2$ oxide phases.

Our results suggest strategies to create atomically-sharp interfaces between crystalline 2D semiconductors and crystalline oxide dielectrics, as desired for the future integration of 2D materials in microelectronics. The dependence of the oxide phase on processing that we show here also suggests that there may be opportunities in engineering the oxide phases that form. We anticipate that this may enable us to adapt the strategies shown here to take advantage of the diversity of hafnium oxides for devices such as resistive switches.

## 5. Methods

### 5.1. Exfoliation and transfer of graphene substrate

We exfoliate few-layer graphene from natural bulk graphite (NGS Trading & Consulting GmbH, Germany) onto SiO$_2$ (90 nm)/Si substrates using Magic Scotch tape (3M). We then use a cellulose acetate butyrate (CAB)-assisted method to transfer the exfoliated graphene flakes to TEM grids.[15,43] To locate the graphene flakes, we used Location-Tagged Holey Nitride TEM Grids with 200 nm thick, 0.5×0.5 mm SiN$_x$ windows containing 2 μm diameter holes with spacing 3 μm, supported on a 3 mm diameter silicon frame (Norcada, Canada). We transferred graphene flakes onto the SiN$_x$ membrane and recorded the images at regions of the flakes suspended over the holes. Both the SiO$_2$ (90 nm)/Si substrate and the TEM grid were cleaned by oxygen plasma before use.

### 5.2. Epitaxial deposition of Hf on graphene in UHV

We bake the graphene substrate in UHV (pressure < 10$^{-9}$ Torr) using a Joule-heating holder at 400 °C for 4 h, to remove polymeric residue from the transfer step and contamination picked up in air. We then raise the temperature to 500–600 °C and deposit Hf using electron-beam evaporation in UHV (pressure ≈ $10^{-9}$ Torr), during which the sample is located between 80–90 cm from the source. The deposition rate is set at 50 Hz min$^{-1}$ (corresponding to ~0.05 nm min$^{-1}$) and a total of 1000 Hz is deposited, where Hz refers to the frequency shift of a quartz crystal microbalance (QCM). For the RT deposition, a total of 2000 Hz is deposited.

### *5.3. Oxidation*

After the Hf metal deposition, we inspect samples in the STEM after a total air exposure time of 15 minutes. We then stored samples in a pumped desiccator for variable lengths of time. We refer to this composite process of air exposure, followed by storage in a desiccator, as "native oxidation". We performed thermal oxidation in air at 400 °C using a box furnace with a heating rate of 30 °C/min, and faster cooling achieved by opening the furnace door.

### *5.4. STEM characterization*

We collected HAADF-STEM images, SAED patterns and EELS of plan view samples using a probe-corrected Thermo Fisher Scientific Themis Z G3 60-300 kV S/TEM operated at 200 kV. We used a probe current of ~100 pA and a semi-convergence angle of 19 mrad for imaging. We prepared cross-sectional samples by FIB using an FEI Helios 600.

### *5.5. Scanning probe measurements*

We performed mid-IR s-SNOM experiments at room-temperature using a s-SNOM system (Neaspec GmbH) equipped with continuous wave, mid-IR quantum cascade lasers (DRS Daylight Solutions Inc.), using an illumination wavelength of 10 μm. We used a Pt-coated silicon tip with a radius of 20 nm, operating in tapping mode at $f = 240$ kHz, and measured the scattered light using a HgCdTe detector. Pseudo-heterodyne interferometric detection was used to extract the near-field signal. The background signal is suppressed by demodulation of the near-field signal at the second harmonic of the tapping frequency. The data $S_2(f)$ presented in **Fig. 5** therefore represents the background-suppressed electrical polarizability of the tip-sample system, which increases sub-linearly with sample conductivity.

To determine sample thickness, we use a Bruker Dimension Icon SPM in tapping mode to carry out AFM measurements on sample regions supported by the SiN$_x$ membrane.

*5.6. Modeling and RMD simulations*

To construct the MD model, we obtain the unit cell of hcp-Hf from the Materials Project database.[44] We expand the unit cell to a 16 × 16 × 6 supercell, resulting in a Hf slab containing 3072 Hf atoms. Then, the hexagonal lattice is converted into an orthogonal lattice, yielding a rectangular Hf slab with a thickness of 2.775 nm, which is comparable to the experimentally measured value of ~2.8 nm (corresponding to a nominal thickness of 0.85 nm with 30 % coverage). Next, we construct a 4-layer graphene substrate with AB stacking, with an initial interlayer spacing of 3.4 Å, to support the Hf slab. The 4-layer graphene structure contains 4800 C atoms. While the number of graphene layers in this model (4 layers) is fewer than the 10–20 layers typically used in experiments, this choice enhances computational efficiency without compromising the ability to investigate the role of graphene in stabilizing the metal and metal oxide phases during annealing simulations. In the last step of building the model, we assemble the Hf slab, the 4-layer graphene, and a variable number of $O_2$ molecules (2–3072 $O_2$) in an orthogonal simulation box with periodic boundary conditions (PBC). The initial distance between the Hf slab and the top graphene layer is set to 5 Å, with a ~24 Å gap between the lateral surfaces of the Hf slab at opposite ends of the simulation box, as determined by the PBC. The modeling parameters for 18 cases, with and without a graphene substrate and with varying initial $O_2$ densities, are summarized in **Table S1**.

We perform ReaxFF MD simulations and ReaxFF uniform-acceptance fbMC/MD hybrid simulations on the Amsterdam Modeling Suite (AMS).[45] For the ReaxFF MD simulations, we first perform energy minimization using the conjugate gradient method. Then, we equilibrate each system with varying numbers of $O_2$ molecules at 300 K in the NVT ensemble for 100 ps. The Berendsen thermostat is applied with a temperature damping parameter of 100 fs. Subsequently, we heat the system from 300 K to 900 K with a heating rate of 10 K/ps. To prevent temperature discrepancies across different components of the system, we apply three Berendsen thermostats separately to the Hf slab, 4-layer graphene, and $O_2$ molecules, each with identical damping parameters (100 fs) and identical start and end temperatures (300 K and 900 K). This approach is necessary because, in systems with high $O_2$ densities, using a single thermostat could lead to an extremely high temperature (up to 2000 K) for the Hf slab and a very low temperature (down to 300 K) for the graphene. During the equilibration and heating steps, the Hf/O interactions in the ReaxFF "ffield" file are disabled to prevent chemical reactions between the Hf slab and $O_2$ molecules. In the subsequent annealing step, the Hf/O interactions in the ReaxFF "ffield" file is enabled. We set a target temperature of 900 K in the NVT ensemble and use three Berendsen thermostats with varying damping parameters: 10,000 fs for the Hf slab and $O_2$ molecules, and 100 fs for the 4-layer graphene. The lower damping parameter for graphene is chosen because it is strongly thermally coupled with the heating source. The higher damping values for the Hf slab and $O_2$ molecules are selected to prevent the clustering of gas-phase Hf oxides and to dissipate excess energy from gas/solid surface interactions, thereby simulating a high-vacuum environment. The annealing simulations run for 2 ns. For all above simulations, the time step is set to 0.1 fs.

For the hybrid simulations, we employ the uniform-acceptance fbMC/MD method, as detailed in previous references.[46–48] The simulations run alternatingly between fbMC and MD steps. During the fbMC steps, the ReaxFF optimizations are alternated with fbMC steps to accelerate events such as $O_2$ adsorption and diffusion into the subsurface of the Hf slab, as well as Hf oxide phase transformations associated with an increased concentration of O depositing deeper into the Hf slab. The parameters for the fbMC steps, including the maximum displacement (drmax), the fbMC frequency (imcfrq), the number of fbMC steps (imcstp), and the mass scaling factor (imcroo), are set to 0.1, 10,000, 10,000, and 4, respectively, in the "control" file.[48] In this setup, the AMS-ReaxFF engine runs alternatingly between 10,000 iterations of MD steps and 10,000 iterations of fbMC steps until the total number of MD steps (nmdit) is reached. During the MD steps, the system is re-equilibrated, and the temperature is gradually reduced to the target value of 900 K. We use the NVT ensemble during the MD steps, with three separate Berendsen thermostats assigned to the Hf slab, the 4-layer graphene, and the $O_2$ molecules, each having temperature damping parameters set to 10,000, 100, and 10,000, respectively. The total numbers of MD iterations and fbMC iterations are each 36,000,000. With a timestep size of 0.1 fs, the MD steps correspond to a total equilibration time of 3.6 ns at 900 K in the NVT ensemble.

Further effects, such as direct bonding between the substrate and the film, and reconstruction of the graphene lattice, are beyond the scope of this study because the Hf/C/H/O ReaxFF force field has not been specifically trained for covalent interactions between graphene and Hf-metal or partially/fully oxidized $HfO_x$.[49]

**Acknowledgements**

This work was carried out in part through the use of MIT.nano facilities. The work was funded in part by Semiconductor Research Corporation (SRC) under contract no.

2021-NM-3027. R. K. acknowledges support by the National Science Foundation Graduate Research Fellowship Program under Grant No. 2141064. R. K. was supported in part by the United States Department of Energy under grant number DE-NA0003965. Financial support was provided by the Two Dimensional Crystal Consortium−Materials Innovation Platform (2DCC-MIP) at The Pennsylvania State University (PSU) under NSF cooperative Agreements DMR-1539916 and DMR-2039351. Computations for this research were performed on the PSU's Institute for Cyber Science Advanced Cyber Infrastructure (ICS-ACI).